\documentstyle[11pt]{article}

\catcode`\@=11
\ifcase \@ptsize
\or % mods for 11 pt
\or % mods for 12 pt
\fi
\advance\textheight by \topskip
\catcode`\@=12

\newcommand{\beqn}{\begin{eqnarray}}
\newcommand{\eeqn}{\end{eqnarray}}
\newcommand{\beqns}{\begin{eqnarray*}}
\newcommand{\eeqns}{\end{eqnarray*}}

\newcommand{\rmd}{\mbox{d}}
\newcommand{\rme}{\mbox{e}}
\newcommand{\rmi}{\mbox{i}}

\newcommand{\wt}{\widetilde}

\newcommand{\re}{\mathop{\mbox{Re}}}

\newcommand{\bfit}[1]{\mbox{\protect\boldmath $#1$}}

\catcode`\@=11
\newcommand{\ccite}[1]
{\@ifundefined{b@#1}{\bf ?}{\@nameuse{b@#1}}}
\catcode`\@=12

\begin{document}

\title{Quantum Annihilation of Anti-de Sitter Universe}

\author{
Iver Brevik\\
Division of Applied Mechanics\\
Norwegian University of Science and Technology\\
N--7491 Trondheim, Norway\\
\and
Sergei D.\ Odintsov\\
Tomsk State Pedagogical University\\
634\,041 Tomsk, Russia
} 

\maketitle

 \begin{abstract}
We discuss the role of conformal matter quantum effects (using 
large $N$ anomaly induced effective action) to creation-annihilation 
of an Anti-de Sitter Universe. The arbitrary GUT
with conformally invariant content of fields
is considered. On a
purely gravitational (supersymmetric) AdS background, the quantum 
effects act against an (already existing) AdS Universe. The annihilation 
of such a Universe occurs, what is common for any conformal matter theory.
On a dilaton-gravitational background, where there is dilatonic contribution 
to the induced effective action, the quantum creation of an AdS Universe is possible 
assuming fine-tuning of the dilaton.

 \end{abstract}

1. There is a large interest now in studies related to 
Anti-de Sitter (AdS) backgrounds. This is caused by several reasons.
First of all, via the AdS/CFT correspondence (for a review,see \cite{review})
investigating the classical IIB supergravity on an AdS background 
(after compactification), one can
get answers for the dual (boundary) quantum gauge theory.
Second, the AdS space is an extremely symmetric one (maximum number of Killing
vectors), 
 like Minkowski space.
Moreover, it is a well-known supersymmetric background for supergravity 
theories. For strings the backgrounds with AdS section are often suspected 
to be the exact vacuum state. 
Third, according to some cosmological data the inflationary Universe 
could have the spatial section with negative curvature. Generalizing 
this for a brane world one can speculate on the possibility of 4d AdS 
stage (or regions) in the early Universe.

The important question is: how should one create the AdS regions 
in the early 
Universe? The mechanisms of such creation maybe important for the realization
 of AdS Black Holes presence. In this note we investigate the role of 
matter quantum effects to the Anti-de Sitter Universe. It is well known 
that the inflationary (de Sitter) Universe maybe created completely 
by quantum effects of conformally invariant matter. On the contrary, 
as we will show below, the quantum creation of Anti-de Sitter Universe 
by quantum effects is rather unrealistic. The quantum {\it annihilation} of
such a
 Universe occurs. Only in  presence of  dilaton 
the fine-tuning of the dilaton 
solution may lead to quantum creation of the AdS Universe (an example of 
quantum maximally supersymmetric YM theory conformally coupled to
conformal background supergravity).    

2. Let us start from one form of the metric describing
four-dimensional Anti-de Sitter (AdS$_4$) spacetime,
\beqn
\label{eq:metric0}
\rmd s^2=\rme^{-2\lambda\wt{x}_3}
(\rmd t^2-(\rmd x^1)^2-(\rmd x^2)^2)-(\rmd\wt{x}^3)^2\;,
\eeqn
having a negative effective cosmological constant
$\Lambda=-\lambda^2$. One may present this metric in conformally flat
form via the transformation
\beqn
y=x^3={\rme^{\lambda\wt{x}_3}\over\lambda}\;.
\eeqn
Then
\beqn
\label{eq:metric1}
\rmd s^2=a^2(\rmd t^2-\rmd\bfit{x}^2)
=a^2\eta_{\mu\nu}\,\rmd x^{\mu}\rmd x^{\nu}\;,
\eeqn
with $a=\rme^{-\lambda\wt{x}^3}=1/(\lambda x^3)=1/(\lambda y)$. This
form of the metric is often useful in the study of quantum gauge
theory via the SG dual. 

Imagine now that the early Universe is described by some quantum grand
unified theory (GUT) containing $N_s$ conformal scalars, $N_f$ spinors and $N_v$
vectors.  It is enough to consider only free fields in such a GUT as
radiative corrections are not important for our purposes. Moreover, the
large class of GUTs - asymptotically finite and asymptotically conformal
GUTs (see the book \cite{BOS} for a review) -  maybe presented as a
 collection of free fields at strong curvature 
 (in the early Universe).  Unlike to the de Sitter space the AdS space  
is  supersymmetric background of GUT if it is SUSY. Note that quantum fields on negative 
curvature space have been reviewed in ref.\cite{sergio}.

The quantum GUT under consideration produces the well-known conformal
anomaly (for a review see \cite{Duff94})
\beqn
T=b\left(F+{2\over 3}\,\Box R\right)+b_1G+b_2\,\Box R\;,
\eeqn
where $b$ and $b_1$ are constants,
\beqn
b={N_s+6N_f+12N_v\over 120(4\pi)^2}\;,\qquad
b_1=-{N_s+11N_f+62N_v\over 360(4\pi)^2}\;,
\eeqn
while $F$ is the square of the Weyl tensor,
\beqn
F=R_{\mu\nu\alpha\beta}R^{\mu\nu\alpha\beta}-2R_{\mu\nu}R^{\mu\nu}
+{1\over 3}\,R^2\;,
\eeqn
and $G$ is the Gauss--Bonnet invariant.
Note that the constant
$b_2$ is known to be in general ambiguous, as it may be
changed by a finite renormalization of the gravitational action.
In the following we put $b_2=0$, since this does not influence
the physical consequences. The features of a particular GUT are encoded in
the numerical values of $b$ and $b_1$ as their signs (what is important
for us) do not change.

Using the above conformal anomaly one can easily calculate the
anomaly-induced effective action \cite{ReigertTseytlinetc}.  Having in
mind the applications for quantum induced AdS space we consider a
metric similar to (\ref{eq:metric1}), i.e.\
$g_{\mu\nu}=\rme^{2\sigma(y)}\eta_{\mu\nu}$, where $\eta_{\mu\nu}$
is the Minkowski metric. Then the techniques of
ref.\ \cite{ReigertTseytlinetc} may easily be applied, and the
following anomaly-induced effective action is obtained,
\beqn
W=\int\rmd^4\!x{\left[2b_1\sigma\Box^2\sigma
-2(b+b_1)(\Box\sigma+\eta^{\mu\nu}
(\partial_{\mu}\sigma)(\partial_{\nu}\sigma))^2\right]}\;.
\eeqn
Since $\sigma$ is assumed to depend only on $y$, this expression
may be simplified:
\beqn
W=V_3\int\rmd y{\left[2b_1\sigma\sigma''''
-2(b+b_1)(\sigma''+(\sigma')^2)^2\right]}\;.
\eeqn
Here $\sigma'=\rmd\sigma/\rmd y$. Formally, this expression coincides
with the effective action on a time dependent conformally flat
background, but its physics is of course different. One should also
remember that the total effective action consists of $W$ plus some
conformally invariant functional. In the case of a conformally flat
background, as considered here, this conformally invariant functional
is a non-essential constant. Only if one considers periodicity on some
of the coordinates (say, AdS BH) will this constant become more
important, as a kind of Casimir energy, since it will depend on the
radius of the compact dimension.

In order to take into account the quantum matter effects in the AdS
Universe we should add the anomaly-induced action to the classical
gravitational action
\beqn
S_{\rm cl}=-{1\over \kappa}\int\rmd^4\!x\,\sqrt{-g}\,(R+6\Lambda)
=-{1\over \kappa}\int\rmd^4\!x\,\rme^{4\sigma}
(-6\rme^{-2\sigma}((\sigma')^2+(\sigma''))+6\Lambda)\;.
\eeqn
The sum of classical action and quantum effective action describes 
the dynamics of whole quantum system.

Equations of motion following from the action $S_{\rm cl}+W$ are
\beqn
\label{eq:motiona}
{a''''\over a}-{4a'a'''\over a^2}
-{3(a'')^2\over a^2}
+\left(6-6{b_1\over b}\right){a''(a')^2\over a^3}
+{6b_1(a')^4\over ba^4}
-{a\over 4b\kappa}\left(-12a''-24\Lambda a^3\right)=0\;.
\eeqn

Here prime means derivative with respect to $y$.
One may now look for the special AdS-like solutions of
Eq.~(\ref{eq:motiona}): $a=c/y$.
When there are no quantum corrections, there is a solution with
$c=1/\sqrt{-\Lambda}$, in accordance with the AdS metric.
When the effective cosmological constant $\Lambda=0$, 
Eq.~(\ref{eq:motiona}) reduces to $c^2=b_1\kappa$
(at $a(y)=c/y$). However, $c^2=b_1\kappa$ leads to an imaginary
scale factor $a$, because $b_1<0$. This indicates that
matter corrections alone can not create an Anti-de Sitter
Universe. This is in contrast to the possibility of creation of
a de Sitter Universe by solely matter quantum effects
\cite{Starobinskyetc}. Indeed, when scale factor depends only on time
the sign of curvature (which is positive) is changing. As a result,
$a=c/\eta$ with $c^2=-b_1\kappa$. In other words, there is always a
solution -  in form of a quantum created de Sitter Universe!
On the contrary, the presence of a negative effective
cosmological constant in classical theory
 is a necessary condition for the existence
of an Anti-de Sitter Universe, at least within our scenario.

In the general case, the algebraic equation for $c^2$ becomes
(assuming $\Lambda<0$):
\beqn
\kappa b_1-c^2-\Lambda c^4=0\;,
\eeqn
and it has the solutions:
\beqn
{c_1}^2=-{1\over 2\Lambda}\left(1+\sqrt{1+4\kappa b_1\Lambda}\right)
\eeqn
and
\beqn
{c_2}^2=-{1\over 2\Lambda}\left(1-\sqrt{1+4\kappa b_1\Lambda}\right).
\eeqn
The first solution corresponds to the quantum corrected 
 Anti-de Sitter Universe. Here, starting from some bare
(even very small!) negative cosmological constant, we get
an Anti-de Sitter Universe with a smaller cosmological constant
 due to quantum corrections.
The quantum corrections act against the existing Anti-de Sitter Universe
and make it less stable. This is the mechanism of annihilation of Anti-de 
Sitter Universe.

The second solution corresponds to the imaginary scale factor
 since it has $c^2<0$.

The complete effective action on the solution $a(y)=c/y$ is
\beqn
S_{\rm cl}+W
&\!\!\!=&\!\!\!
V_3\int{\rmd y\over y^4}{\left[
6b_1\ln\!\left({c^2\over y^2}\right)-8(b+b_1)-
{6(\Lambda c^4-c^2)\over\kappa}\right]}
\nonumber\\
&\!\!\!=&\!\!\!
V_3\int{\rmd y\over y^4}{\left[
6b_1\ln\!\left({c^2\over y^2}\right)-8b-14b_1
+{12c^2\over\kappa}\right]}\;.
\eeqn
The dependence of effective action from $c^2$ is seen.

Thus, we have shown that quantum corrections in
an already existing Anti-de Sitter Universe make it less stable,
whereas creation of an Anti-de Sitter Universe by
quantum corrections alone is impossible. This is in contrast to the
 de Sitter Universe, which may be created by
quantum corrections alone.
It is now interesting to understand the role of other effects 
in our scenario. As such we consider the presence of dilaton.
The related dilatonic terms in action may play the role of effective 
cosmological constant.

3. As an explicit example of theory where dilaton appears in conformal
anomaly we consider  ${\cal N}=4$ $\mbox{SU}(N)$ super YM 
theory covariantly coupled with the background ${\cal N}=4$ 
conformal supergravity (see \cite{Kakuetc78} for an introduction). 
The corresponding vector multiplet is $(A_{\mu},\psi_i,X_{ij})$. 
There is also complex scalar (axion and dilaton) in the conformal  
supergravity multiplet. Note that such a theory is not a realistic one. 

On the purely bosonic background the conformal anomaly in such a theory is
the following 
 \ \cite{LiuTseytlin98}

\beqn
\label{eq:conformalanomaly}
T=b\left(F+{2\over 3}\,\Box R\right)+b_1G+b_2\Box R
+C{\left[\Box\phi^{\ast}\Box\phi
-2\left(R^{\mu\nu}-{1\over 3}\,g^{\mu\nu}\,R\right)
\partial_{\mu}\phi^{\ast}\,\partial_{\nu}\phi\right]}.
\eeqn
The last term, with the constant
$$
C={N^2-1\over(4\pi)^2}\;,
$$
is the contribution of dilaton and axion fields.
Note also that in
adjoint representation $N_v=N^2-1$, $N_s=6N_v$ and  
$N_f=2N_v$ in the theory under discussion. 

Using the conformal anomaly (\ref{eq:conformalanomaly}), and following
ref.\ \cite{BrevikOdintsov99}, it is again not difficult to construct
the anomaly-induced effective action $W$ on the conformally flat
background, $g_{\mu\nu}=\rme^{2\sigma}\,\eta_{\mu\nu}$.

With $\sigma$ and $\phi$ depending only on time, one gets,
in terms of the conformal time $\eta$ \cite{BrevikOdintsov99},
\beqn
W=V_3\int\rmd\eta{\left[2b_1\sigma\sigma''''
-2(b+b_1)(\sigma''+(\sigma')^2)
+C\sigma\re(\phi^{\ast}\phi'''')\right]},
\eeqn
where $V_3$ is three-dimensional volume, and
$\sigma'=\rmd\sigma/\rmd\eta$. As usual, the complete effective
action is the sum of two terms: $W$ and some conformally invariant
functional $W_1$. Since we discuss a conformally flat background,
$W_1$ can only depend on the complex scalar $\phi$, and has to be
constant when $\phi$ is constant. Supposing that $W_1$ is a local
functional (so that the Schwinger--De Witt expansion may be used),
we are left with only one possibility,
\beqn
W_1=V_3\int\rmd\eta\,\re(\phi^{\ast}\phi'''')\ln\mu^2\;.
\eeqn
The coefficient of $W_1$ depends on the regularization used,
since it may be absorbed into the scale $\mu$, by a redefinition.
Thus our quantum correction to the classical action is
$\Gamma=W+W_1$.

The usual choice for the complex scalar $\phi$ is
\beqn
\phi=\chi+\rmi\rme^{-\varphi}\;,
\eeqn
where $\varphi$ is the dilaton, and $\chi$ is the R-R scalar (axion)
of type IIB supergravity (or conformal supergravity, as above).

Thus, the final expression for the one-loop effective action of
${\cal N}=4$ super Yang--Mills theory coupled to conformal 
supergravity on a conformally flat
background is
\beqn
\Gamma=V_3\int\rmd\eta{\left[2b_1\sigma\sigma''''
-2(b+b_1)(\sigma''+(\sigma')^2)
+(C\sigma+A)\re(\phi^{\ast}\phi'''')\right]}\;,
\eeqn
where $A$ is some constant depending on regularization, and
$\phi=\chi+\rmi\rme^{-\varphi}$. One may consider the case of
large $N$, then in the large $N$ expansion the 
proper quantum gravity corrections to $\Gamma$
will be of next-to-leading order.

There are different choices for the classical gravitational action.
For example, one can consider
 the axion-dilatonic gravity by
Gibbons--Green--Perry \cite{Gibbonsetal96}, which describes
the bosonic sector of type IIB supergravity,
\beqn
\label{eq:Scldilax}
S_{\rm cl}=-{1\over\kappa}\int\rmd^4\!x\,\sqrt{-g}\,\left(R
+{1\over 2}\,g^{\mu\nu}\,
\partial_{\mu}\varphi\,\partial_{\mu}\varphi
+{1\over 2}\,\rme^{2\varphi}\,g^{\mu\nu}\,
\partial_{\mu}\chi\,\partial_{\mu}\chi\right).
\eeqn
In the absence of the dilaton and axion, this reduces to
the standard action of general relativity. Using the above
choice of a conformally flat metric one may simplify
Eq.~(\ref{eq:Scldilax}) as follows,
\beqn
\label{eq:Scldilaxsimpl}
S_{\rm cl}=-{1\over\kappa}\,V_3\int\rmd\eta\left(
6\rme^{2\sigma}\,((\sigma')^2+(\sigma''))
+{1\over 2}\,\rme^{2\sigma}\,(\varphi')^2
+{1\over 2}\,\rme^{2\sigma+2\varphi}\,(\chi')^2\right).
\eeqn

Now it is convenient to transform to cosmological time $t$,
such that $\rmd t=a(\eta)\,\rmd\eta=\rme^{\sigma(\eta)}\,\rmd\eta$.
The equations of motion may be obtained by  variations with respect to
$a$, $\varphi$ and $\chi$.

We will consider the simpler choice when the axion is equal to zero, and
the kinetic term for the dilaton in the classical action is absent. The 
regularization where $A=0$ is also chosen. As a result one comes 
to the model discussed in ref.\cite{BrevikOdintsov99}. The study 
of the corresponding effective equations for time-dependent conformally 
flat metric has been done in the above work. The possibility of quantum creation of
de Sitter Universe has been proved.
One can give now an analysis related to a quantum corrected Anti-de Sitter 
Universe in the same way as it was done in previous section. 
Note, however, that in the case under discussion there is no classical
cosmological constant and as a result there is no Anti-de Sitter solution 
on the classical level.

The effective equations of motion take the form:
\beqn
\label{eq:motionavarphi}
{a''''\over a}-{4a'a'''\over a^2}-{3a''^2\over a^2}
+{6a''a'^2\over a^3}\left(1-{b_1\over b}\right)
+{6b_1a'^4\over ba^4}+{3aa''\over\kappa b}
-{C\over 4b}\,\varphi\varphi''''&\!\!\!=&\!\!\!0\;,
\nonumber\\
\ln a\;\varphi''''+(\ln a\,\varphi)''''&\!\!\!=&\!\!\!0\;.
\eeqn
Here prime means derivative on $y$ and we put $\varphi=\phi$.

Motivated by ref.~\cite{BrevikOdintsov99} one can make now
the following transformation:
\beqn
\rmd z=a(y)\,\rmd y\;.
\eeqn
Then, in terms of the variable $z$ the first of
equations~(\ref{eq:motionavarphi}) is:
\beqn
\label{eq:motionaII}
a^2\,\mathop{a}^{....}+3a \dot{a}\,\mathop{a}^{...}
+a\ddot{a}^2-\left(5+{6b_1\over b}\right)\dot{a}^2\ddot{a}
+{3\over\kappa b}\left(a^2\ddot{a}+a\dot{a}^2\right)
-{C\varphi\,Y{[\varphi,a]}\over 4b}=0\;.
\eeqn
Here $\dot{a}=\rmd a/\rmd z$, $Y{[\varphi,a]}$ is given
in ref.~\cite{BrevikOdintsov99}, and the second
equation~(\ref{eq:motionavarphi}) (in terms of $z$) is also given
in ref.~\cite{BrevikOdintsov99} (Eq.~(10)).
The only difference from the analysis done there, is that
the sign of the $1/\kappa$ term is reversed. Then,
as in ref.~\cite{BrevikOdintsov99} we search for special
solutions
\beqn
a(z)\simeq a_0\,\rme^{Hz}\;,\qquad
\varphi(z)\simeq\varphi_0\,\rme^{-\alpha Hz}\;,
\eeqn
Analyzing the second of equations~(\ref{eq:motionavarphi})
(in terms of $z$) and dropping logarithmic term in it
(same arguments as in ref.~\cite{BrevikOdintsov99} may be given)
one comes to the same solution:
\beqn
\varphi(z)
=\varphi_1\,\rme^{-{3\over 2}\,Hz}+
\varphi_2\,\rme^{-2.62\,Hz}
+\varphi_3\,\rme^{-0.38\,Hz}\;,
\eeqn
where $\varphi_1,\varphi_2,\varphi_3$ are constants.
Substituting the particular solution
$\varphi(z)=\varphi_0\,\rme^{-\alpha Hz}$
in Eq.~(\ref{eq:motionaII}) one obtains:
\beqn
H^2\simeq{1\over\kappa}{\left[b_1+{C\over 24}\,{\varphi_0}^2
\left(\alpha^4-6\alpha^3+11\alpha^2-6\alpha\right)\right]}^{-1}\;.
\eeqn
The first term in the denominator is always negative, while
the second term may be positive only at $\alpha=3/2$.
Then, it is only with the special dilaton solution
$\varphi(z)=\varphi_1\,\rme^{-{3\over 2}\,Hz}$ 
(i.e.\ $\varphi_2=\varphi_3=0$) and at the condition
${\varphi_1}^2>12$ one gets the positive $H^2$ and,
hence, non-imaginary scale factor for AdS Universe.
Thus, there occurs the possibility for quantum creation of
dilatonic AdS Universe, but it requires a strong fine-tuning
of the dilatonic solution. This is, of course, a very non-realistic
situation. Note that the corresponding AdS scale factor is:
\beqn
a(y)=-{1\over H y}\;.
\eeqn
One can analyze the Eqs.~(\ref{eq:motionavarphi}) numerically
with the same qualitative result; for some very special
initial conditions for dilaton and scale factor the quantum
creation of AdS Universe is possible. However, in most cases
such a process does not occur.

Thus, we demonstrated that quantum effects of conformal matter 
do not support the creation of an AdS Universe. Moreover, an already 
existing classical AdS Universe is annihilating due to quantum 
effects. Only for dilaton-gravitational background (maximally 
SUSY YM theory) where there is dilatonic contribution to 
conformal anomaly and to induced effective action the quantum 
effects may create AdS Universe subject to strong fine-tuning. 
However, the probability of such creation is much less than 
the same process for de Sitter Universe at similar conditions.

The model under discussion maybe understood also as a simplified 
model for creation-annihilation of AdS Black Hole. Taking additional 
contribution to effective action due to geometrical structure of 
AdS BH one can repeat this analysis for a more realistic situation.
Unfortunately, this additional piece of effective action is not 
known in closed form.
 From another side, the inclusion of non-trivial axion maybe 
also done. However, in such case only numerical analysis of
equations of motion is possible.

Acknowledgments. The work by SDO has been supported in part 
by the Norwegian Research Council and RFBR grant N99-02-16617.
 We are extremely grateful to
Jan Myrheim for participation at the early stage of this work and 
helping in preparation of draft of this ms. We also thank K{\aa}re Olaussen
 and Shinichi  Nojiri for
useful discussions.

\end{document}